# Anharmonic Waves in Field Theory

F. J. Himpsel

Physics Department, University of Wisconsin Madison, Madison WI 53706, USA

**Abstract**

This work starts from the premise that sinusoidal plane waves cease to be solutions of field theories when turning on an interaction. A nonlinear interaction term generates harmonics analogous to those observed in nonlinear optical media. This calls for a generalization to anharmonic waves in both classical and quantum field theory. Three simple requirements make anharmonic waves compatible with relativistic field theory and quantum physics. Some non-essential concepts have to be abandoned, such as orthogonality, the superposition principle, and the existence of single-particle energy eigenstates. The most general class of anharmonic waves allows for a zero frequency term in the Fourier series, which corresponds to a quantum field with a non-zero vacuum expectation value. Anharmonic quantum fields are defined by generalizing the expansion of a field operator into creation and annihilation operators. This method provides a framework for handling exact quantum fields, which define exact single particle states.

───────────────────────
Electronic address: fhimpsel@wisc.edu

**Contents**

# 1. General Concept

Nonlinear wave equations and field theories have attracted considerable interest for more than a century [1]-[22]. A nonlinear term describes interactions. Otherwise the superposition principle would hold, i.e., two waves would penetrate each other without creating a scattered wave. The nonlinearity is frequently hidden by the coupling of two different fields, such as the Maxwell and Dirac fields in quantum electrodynamics. It reveals itself after eliminating the Maxwell field via its Green's function [14]. Explicit nonlinear modifications of a single field are more transparent and have been used for Maxwell's equations [1],[2], for the Dirac equation (see [3] and references therein), and for the Klein-Gordon equation [4-6]. Einstein's equations for the gravitational field are intrinsically nonlinear [7,8]. Since nonlinear interactions are essential for a realistic description of nature, a possible effect of nonlinearity on quantum mechanics itself has been considered [9]-[13].

Plane wave solutions of field equations play a particularly important role, because they preserve the homogeneity of space-time. Sinusoidal plane waves, in particular, serve as basis set for the Fourier expansion of an arbitrary solution in classical field theory. They are used in quantum field theory for the expansion of field operators into creation and annihilation operators of particles with well-defined energy and momentum. Since they are solutions of the non-interacting field equations, sinusoidal plane waves are essential for perturbation theory. However, they need to be generalized in order to obtain solutions of interacting field theories. Exact solutions have been elusive for realistic quantum field theories, such as quantum electrodynamics and the other components of the Standard Model.

Such arguments provide strong motivation to explore periodic plane wave solutions in the presence of nonlinear interactions. They have been explored not only in field theory [1]-[8], but also in the soliton literature [15]-[22]. Although a soliton by itself is not periodic, it is sometimes possible to construct a periodic solution from a train of regularly-spaced solitons [19]-[21]. For the solutions of such nonlinear equations one needs to consider generalized versions of sine and cosine [22], such as Jacobi's elliptic functions [19],[21].

To get an intuitive picture of nonlinear effects, consider an intense laser beam propagating in a nonlinear optical medium [23]. The light wave loses its sinusoidal character and develops harmonics. These match the temporal and spatial periodicity of the fundamental wave, as required by the homogeneity of space-time. An inversion-symmetric optical medium creates harmonics separated by twice the fundamental frequency ($3^{rd}$, $5^{th}$, etc.). When inversion symmetry is broken, the harmonics become single-spaced, starting with the $2^{nd}$. Since higher harmonics tend to weaken rapidly, the lowest harmonic already provide substantial insight into nonlinear phenomena. This is particularly true for field theories with small coupling constants, such as quantum electrodynamics.

Before venturing into quantum field theory, classical fields will be used as testing ground for investigating the effect of nonlinearity. These correspond to single-particle wave functions. They are closely related to quantum field operators, which will be defined in Section 7 by an expansion into single-particle wave functions combined with creation and annihilation operators of single-particle states. Rather than looking for the solutions of specific wave equation, a top-down approach will be taken. The most general class of anharmonic waves will be sought out which satisfies the requirements of relativistic field theory and quantum physics. These include relativistic invariance, homogeneity of space-time, completeness, probability conservation, gauge invariance, and so on. Anharmonic waves then will be parametrized by two types of Fourier series, which may serve as *ansatz* for constructing exact plane wave solutions.

Three simple constraints are sufficient to satisfy all the necessary criteria. They leave room for a large variety of wave functions, which can be categorized into three symmetry classes. The lowest possible symmetry class contains a $-1^{st}$ harmonic, which corresponds to a constant. That leads to a non-zero vacuum expectation value in quantum field theory, a hallmark of spontaneous symmetry breaking and the generation of mass.

The move from classical fields to quantum fields is equivalent to replacing a single-particle wave function by many-body theory. Technically this is achieved by an expansion of quantum field operators into creation and annihilation operators multiplied by plane waves (see Section 7). This expansion can easily be generalized by substituting anharmonic waves for sinusoidal waves.

## 2. Criteria for Anharmonic Waves

The criteria for the most general class of anharmonic waves can be condensed into three simple conditions for a scalar wave:

(1) $\psi(z+2\pi) = \psi(z)$

(2) $\psi(z)^* \cdot \psi(z) = 1$

(3) $\psi(-z)^* = \psi(z)$

The boundary conditions $\psi(0) = 1$, $\psi(\pi) = -1$ are the same as for the sinusoidal wave $\exp(iz)$. The covariant variable $z = -p_\mu x^\mu = (\mathbf{p}\,\mathbf{x} - E\,t)$ describes a plane wave. For spinor, vector, and tensor waves one needs to multiply the scalar wave by a pre-factor that does not vary in space-time, but depends on the momentum $\mathbf{p}$, such as the spinors $u(\mathbf{p},s)$, $v(\mathbf{p},s)$ for electrons and positrons, the polarization vectors $\varepsilon^i_\mu(\mathbf{k})$ for photons, and 4×4 tensors for gravity waves [8].

Conditions (1)-(3) make sure that basic symmetries of quantum physics and relativity are satisfied. Condition (1) defines anharmonic waves as periodic with the same period as the sinusoidal waves $\exp(iz)$. Condition (2) determines the normalization, again the same as for $\exp(iz)$. More importantly, it makes sure that the probability density is constant and thereby preserves the homogeneity of space-time. Condition (3) is connected to CPT symmetry, since charge conjugation (C) substitutes $\psi \rightarrow \psi^*$ and space-time inversion (P,T) substitutes $x^\mu \rightarrow -x^\mu$. Their combination reproduces the wave function according to (3). CPT symmetry can also be interpreted as the identification of an antiparticle as a missing particle with negative energy and momentum (a hole). In that case one inverts $p_\mu \rightarrow -p_\mu$ instead of $x^\mu$. Anharmonic waves automatically preserve gauge symmetry, since gauge invariance depends only on the form of the Lagrangian, not on specific solutions of the Euler-Lagrange equations. Anharmonic waves form a complete basis set, because every sinusoidal wave can be expanded into anharmonic waves (see Section 6). Orthogonality, however, is not preserved. Two waves whose momenta that are connected by a rational factor share common harmonics and thus have a finite overlap integral. Completeness is necessary for a basis set, but orthogonality is not. Another victim of reduced symmetry is the superposition principle. This is hardly surprising, since interacting fields have nonlinear terms that violate the superposition principle. Two interacting waves create additional scattered waves.

When cosine and sine are generalized to anharmonic waves, some of their symmetry properties are at stake. For example, the real and imaginary parts of ψ may have different shapes, as shown in Fig. 1. In this example the real part of ψ is "bottom-heavy" (blue curve) and the imaginary part "top-heavy" (red). The top-heavy version can be viewed as a sinusoidal wave that has gone through a nonlinear amplifier whose gain is reduced at large amplitudes. For the bottom-heavy version the gain increases with the amplitude.

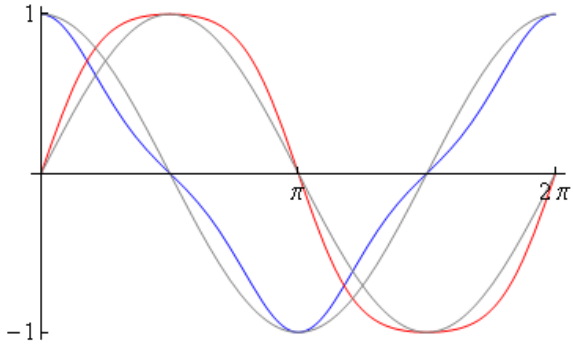

**Figure 1** Real and imaginary part of a Class 2 anharmonic wave (blue and red). Cosine and sine are shown in gray. The real part is "bottom-heavy" and the imaginary part "top-heavy". Systematic plots for various symmetry classes can be found in Appendix C, Fig. 4. From Equations (5a,b) with $\varphi_2(z) = \frac{1}{4}\sin(2z)$.

### 3. Fourier Series of Anharmonic Waves

Condition (1) implies that ψ(z) can be expanded into a Fourier series:

(4a) $$\psi_1(z) = \sum_{m=-\infty}^{\infty} a_m \cdot \exp[i(m+1)z] \qquad \text{Class 1}$$

$$\psi_2(z) = \sum_{m=-\infty}^{\infty} a_m \cdot \exp[i(2m+1)z] \qquad \text{Class 2}$$

$$\psi_3(z) = \sum_{m=-\infty}^{\infty} a_m \cdot \exp[i(4m+1)z] \qquad \text{Class 3}$$

These three Fourier series define three symmetry classes. Class 1 contains all harmonics, Class 2 only harmonics spaced by even numbers, and Class 3 only harmonics spaced by multiples of 4. Their properties will be investigated in Section 4 and Appendices C, D, E. Condition (2) imposes an infinite number of constraints on the Fourier coefficients $a_m$, as discussed in Appendix A. The most important constraint has the form of a sum rule:

(4b) $$\sum_{m=-\infty}^{\infty} a_m^2 = 1 \qquad a_m \text{ real}$$

The infinite number of remaining constraints can only be satisfied by an infinite Fourier series of the type (4a). However, one can find a different Fourier representation that satisfies all three conditions (1)-(3) without extra constraints. This is achieved by introducing a periodic phase shift $\varphi(z)$ and expanding it into a Fourier series:

(5a) $\quad \psi(z) = \exp[i(z + \varphi(z))]$

(5b) $\quad \varphi_1(z) = 2 \cdot \sum_{m=1}^{\infty} c_m \cdot \sin(m\,z) \qquad$ Class 1

$\qquad \varphi_2(z) = 2 \cdot \sum_{m=1}^{\infty} c_m \cdot \sin(2m\,z) \qquad$ Class 2

$\qquad \varphi_3(z) = 2 \cdot \sum_{m=1}^{\infty} c_m \cdot \sin(4m\,z) \qquad$ Class 3

Figure 2 illustrates how such a phase shift creates a Class 2 anharmonic wave.

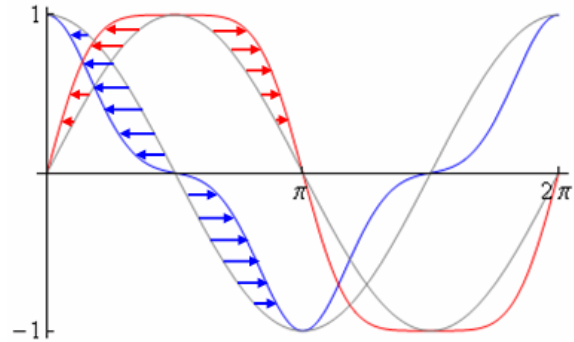

**Figure 2** Creation of anharmonic waves via a periodic modulation of the phase $\varphi(z)$ in (5a,b), with the real part in blue and the imaginary part in red. The phase shift is indicated by arrows. This construction satisfies $\psi(z)^* \cdot \psi(z) = 1$ automatically. Shown for a Class 2 wave with $\varphi_2(z) = 0.45 \sin(2z)$.

The periodic phase $\exp[i\varphi(z)]$ in (5a) resembles a U(1) gauge transformation of the wave function $\psi(z)$. However, there is no corresponding gauge boson $A_\mu$ that can be transformed together with $\psi$ to maintain gauge symmetry. Therefore, anharmonic waves are not equivalent to sinusoidal waves via U(1) gauge transformations.

The Fourier series of the phase involves only sine functions because of condition (3). This can be seen by splitting (3) into its real and imaginary parts. The real part has to be even in z, and the imaginary part odd, as one would expect from a generalized version of cos and sin. In order for $\sin[(z+\varphi(z)]$ to be odd in z, its argument has to be odd, and that implies that $\varphi(z)$ is odd in z.

The factor of 2 in (5b) has been introduced in order to have a simple relation between the leading Fourier coefficients $a_{\pm 1}$ and $c_1$ for small anharmonicity:

(5c) $\boxed{a_{\pm 1} \approx \pm c_1}$

This can be seen by expanding $\exp[i\,\varphi(z)]$ into a Taylor series and neglecting harmonics with m>1. For Class 1 this expansion takes the form:

$$\exp[i\,\varphi_1(z)] \approx 1 + i\,\varphi_1(z) \approx 1 + i\,2\sum c_m \cdot \sin(m z) \approx 1 + i\,2\,c_1 \cdot \sin(m z)$$
$$= 1 + c_1 \cdot \exp(+i z) - c_1 \cdot \exp(-i z)$$

A similar expansion confirms (5c) for Class 2 and 3. This relation shows that negative harmonics cannot be avoided, since the coefficients $a_{\pm 1}$ for the dominant harmonics are comparable in magnitude. This is different from nonlinear optics, where negative harmonics do not exist.

## 4. Symmetry Classes

The three classes of anharmonic waves in (4a) and (5b) exhibit distinct symmetry properties. These are summarized in Table 1 and illustrated in Fig. 3a (Class 1), Fig. 1 (Class 2), Fig. 3b (Class 3), and more systematically in Fig. 4 (Appendix C).

**Table 1** Properties of the three symmetry classes of anharmonic waves defined in (4a), (5b). A class with higher number includes all the symmetries listed for the lower classes.

Class 1: $\quad \psi_1(\pm\pi) = -1 \quad\quad \psi_1(z+2\pi) = \psi_1(z)$

Class 2: $\quad \psi_2(\pm\pi/2) = \pm i \quad\quad \psi_2(z+\pi) = -\psi_2(z)$

Class 3: $\quad \psi_3(\pm\pi/4) = (1\pm i)/\sqrt{2} \quad\quad \mathrm{Im}[\psi_3(z+\pi/2)] = \mathrm{Re}[\psi_3(z)]$

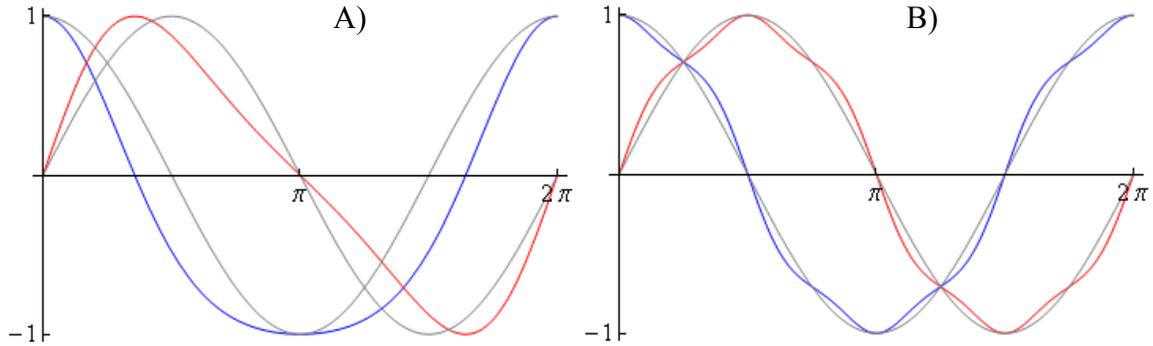

**Figure 3** Real and imaginary part (blue and red) of anharmonic waves in A) Class 1 and B) Class 3. Cosine and sine are shown in gray. The real part of Class 1 has a non-zero average which gives rise to a non-zero vacuum expectation value in quantum field theory. From (5a,b) with $\varphi_1(z) = \frac{1}{2}\sin(z)$ for A) and $\varphi_3(z) = \frac{1}{8}\sin(4z)$ for B).

For each of the three classes one can distinguish a sub-set with an extra symmetry:

(6) $\boxed{\begin{array}{l} a_{-m} = (-)^m \cdot a_m \\ c_{2m} = 0 \end{array}}$ $\quad\quad \begin{array}{l} m = -\infty, \ldots, \infty \\ m = 1, \ldots, \infty \end{array}$

The conditions for the $a_m$ and $c_m$ are equivalent. This extra symmetry gives equal strength to positive and negative harmonics. It is not apparent from the shape of the wave functions, however, since it does not affect the properties listed in Table 1.

Class 1 represents the most general Fourier series, which may contain coefficients $a_m$ with both odd and even m. As shown in Figure 3A, the real part of $\psi$ oscillates asymmetrically, which leads to a constant Fourier term $a_{-1} \cdot \exp(0 \cdot i\, z)$. It produces a non-zero average of a classical field and a non-zero vacuum expectation value of a quantum field. This property might become useful for theories with spontaneous symmetry breaking. The term $a_{+1}$ corresponds to frequency doubling which occurs in nonlinear laser optics when the inversion symmetry is broken, for example at surfaces and in crystals that lack inversion symmetry. Class 1 waves should be relevant to interactions with spontaneous symmetry breaking by a non-zero vacuum expectation value, such as the weak interaction.

Class 2 anharmonic waves contain exclusively odd harmonics $a_m$. Such waves are shown in Figures 1 and 2. Although they exhibit symmetry about the quarter-period points ½π, the real and imaginary part remain dissimilar. One of them is "top-heavy", the other "bottom-heavy", like the distortion of a sine wave produced by an amplifier whose gain decreases or increases at large amplitude. Symmetry can be restored to some degree, because "top-heavy" switches with "bottom-heavy" when changing the sign of the asymmetry parameter q defined in Appendix A (see Fig. 4 in Appendix C). Noteworthy examples of Class 2 functions are Jacobi's elliptic functions, which are periodic along both the real and imaginary axis (see Appendix D). Class 2 waves are candidates for interactions without symmetry breaking, such as the electromagnetic interaction in quantum electrodynamics.

Class 3 contains only Fourier coefficients $a_m$ spaced by multiples of 4. Real and imaginary part have the same shape (Figure 3B), like for cosine and sine. However, this symmetry requirement seems excessive, since it introduces extra oscillations of the anharmonic wave which increase the kinetic energy. Two types of line shapes are possible which may be characterized as "bottom-heavy at the top" or "top-heavy at the top". Both are shown in Fig. 4 (Appendix C).

The free parameters available to construct anharmonic waves are the coefficients $c_m$ of the Fourier expansion (5). They are not constrained, while the coefficients $a_m$ of the regular Fourier series (4) require an infinite number of constraints to satisfy (2). These are given in Appendix A. This explains why there roughly twice as many $c_m$ as $a_m$, since the $c_m$ range from 1 to ∞ compared to −∞ to ∞ for the $a_m$.

In practice, one can expect only a few significant Fourier coefficients for a field theory with a small coupling constant, such as quantum electrodynamics. Most of the nonlinear effects should be captured by the lowest harmonic $a_{\pm 1} \approx \pm c_1$. This leads to an approximation containing only the anharmonicity parameter q.

In the following we will focus on Class 2 anharmonic waves. They are the most likely candidates for highly-symmetric quantum field theories, such as quantum electrodynamics. Analogous results can be obtained for Class 1 and 3.

## 5. Energy and Momentum

Energy E and momentum **p** of the higher harmonics scale like (2m+1) for Class 2 anharmonic waves due to their covariant definition in (4a) as a function of $(2m+1)z = ((2m+1)\mathbf{p}\mathbf{x} - (2m+1)Et)$. This ensures that the phase velocity $v_{ph}$ of all harmonics is the same as that of the fundamental, such that they remain synchronized:

(7) $\quad v_{ph} = \dfrac{(2m+1)\,E}{(2m+1)\,|\mathbf{p}|} = \dfrac{E}{|\mathbf{p}|} \qquad\qquad E = (\mathbf{p}^2 + M^2)^{½}$

Another consequence of the scaling factor (2m+1) is an increase in the effective mass $M_{eff}$ for the harmonics:

(8) $\quad M_{eff} = \left( [(2m+1)E]^2 - [(2m+1)\mathbf{p}]^2 \right)^{½} = |2m+1| \cdot M$

However, the energy and momentum operators $E = i\,\partial/\partial t$ and $\mathbf{p} = -i\,\partial/\partial \mathbf{x} = -i\,\nabla$ do not produce eigenvalues for energy and momentum anymore. This can be seen most clearly from the Fourier expansion of the phase in (5a,b). With $z = (\mathbf{p}\mathbf{x} - Et)$ one obtains:

(9) $\quad i\,\partial/\partial t\ \psi_2(z) = E \cdot \psi_2(z) \cdot (1 + [2 \sum 2m\, c_m \cos(2mz)])$

$\quad\quad -i\,\nabla\ \psi_2(z) = \mathbf{p} \cdot \psi_2(z) \cdot (1 + [2 \sum 2m\, c_m \cos(2mz)])$

The harmonics produce extra oscillatory terms (in square brackets), which prevent the energy-momentum operator from reproducing an anharmonic wave. The role of these terms becomes clearer after multiplying (9) with $\psi_2^*$ from the left to determine the energy-momentum density. Utilizing (2) $\psi_2^* \psi_2 = 1$ one obtains the normal, constant energy-momentum density plus propagating energy density waves of the form $\cos(2mz)$. These move at the same phase velocity as the fundamental: $v_{ph} = (2m \cdot E)/(2m \cdot |\mathbf{p}|) = E/|\mathbf{p}|$. They oscillate with the difference frequency between the harmonics and the fundamental.

Although E and **p** are not eigenvalues anymore, their expectation values remain unchanged, since the average over the oscillatory term $\cos(2mz)$ vanishes:

(10) $\langle i\partial/\partial t \rangle = \langle \psi_2(z)^* (i\,\partial/\partial t)\, \psi_2(z) \rangle = E \cdot \langle\, 1 + [2 \sum 2m\, c_m \cos(2mz)] \,\rangle = E$

$\langle -i\nabla \rangle = \langle \psi_2(z)^* (-i\nabla)\, \psi_2(z) \rangle = \mathbf{p} \cdot \langle\, 1 + [2 \sum 2m\, c_m \cos(2mz)] \,\rangle = \mathbf{p}$

To calculate the energy uncertainty $\delta E = (\langle (i\partial/\partial t)^2 \rangle - \langle i\partial/\partial t \rangle^2)^{1/2}$ and the momentum uncertainty $\delta p = (\langle (-i\nabla)^2 \rangle - \langle -i\nabla \rangle^2)^{1/2}$ we need the following terms:

(11) $\langle \psi_2(z)^* \cdot (i\partial/\partial t)^2\, \psi_2(z) \rangle = E^2 \cdot (1 + \langle [2 \sum (2m)\, c_m \cos(2mz)]^2 \rangle)$

$\langle \psi_2(z)^* \cdot (-i\nabla)^2\, \psi_2(z) \rangle = \mathbf{p}^2 \cdot (1 + \langle [2 \sum (2m)\, c_m \cos(2mz)]^2 \rangle)$

The resulting energy and momentum uncertainties are:

(12) $\delta E = E \cdot \langle [2 \sum (2m)\, c_m \cos(2mz)]^2 \rangle^{1/2} \approx E \cdot \sqrt{8}\,|c_1|$

$\delta p = |\mathbf{p}| \cdot \langle [2 \sum (2m)\, c_m \cos(2mz)]^2 \rangle^{1/2} \approx |\mathbf{p}| \cdot \sqrt{8}\,|c_1|$

The uncertainty $\delta E$ reflects the possibility for energy to move back and forth between harmonics and the fundamental. Similar oscillations have been observed by quantum beat spectroscopy of narrowly-spaced energy levels under coherent excitation [24]. There is also a loose connection with oscillations between different mass eigenstates of neutrinos.

The question arises whether a more general definition of the energy and momentum operators can be found that has anharmonic waves as eigenstates. After all, anharmonic waves preserve translation invariance in time and space, which is responsible for the conservation of energy and momentum. However, linear operators and their eigenvalues are mismatched with non-linear wave equations, where the amplitude of an eigenfunction cannot be chosen arbitrarily. Furthermore, single-particle properties are only approximate in an interacting many-body theory. Only the total energy and momentum should remain conserved, as defined via the energy-momentum tensor.

## 6. Orthonormality and Completeness

The Fourier series (4a) provides a convenient way to test whether the normalized anharmonic waves $(2\pi)^{-3/2}\psi_2(z)$ have the same orthonormality and completeness properties as sinusoidal waves:

(13) $\quad (2\pi)^{-3} \int \exp(i\,\mathbf{p}\,\mathbf{x})^* \cdot \exp(i\,\mathbf{p}'\mathbf{x})\ d^3x\ =\ \delta^3(\mathbf{p}-\mathbf{p}')$ $\qquad$ orthonormality

(14) $\quad (2\pi)^{-3} \int \exp(i\,\mathbf{p}\,\mathbf{x})^* \cdot \exp(i\,\mathbf{p}\,\mathbf{x}')\ d^3p\ =\ \delta^3(\mathbf{x}-\mathbf{x}')$ $\qquad$ completeness

For anharmonic waves of Class 2 these relations take the following form:

(15) $\quad (2\pi)^{-3} \int \psi_2(\mathbf{p}\,\mathbf{x})^* \cdot \psi_2(\mathbf{p}'\mathbf{x})\ d^3x\ =\ \sum_{m,m'} a_m\, a_{m'} \cdot \delta^3\!\left((2m+1)\mathbf{p} - (2m'+1)\mathbf{p}'\right)$

(16) $\quad (2\pi)^{-3} \int \psi_2(\mathbf{p}\,\mathbf{x})^* \cdot \psi_2(\mathbf{p}\,\mathbf{x}')\ d^3p\ =\ \sum_{m,m'} a_m\, a_{m'} \cdot \delta^3\!\left((2m+1)\mathbf{x} - (2m'+1)\mathbf{x}'\right)$

The term $m=m'=0$ approximately replicates the $\delta$-functions of the orthonormality and completeness relations for sinusoidal waves (not exactly because of $a_0 < 1$). However, there are numerous additional $\delta$-functions. Orthogonality is spoiled whenever the momenta $\mathbf{p}$ and $\mathbf{p}'$ in (15) satisfy the relation $(2m+1)\mathbf{p}=(2m'+1)\mathbf{p}'$, i.e. when $\mathbf{p}'$ and $\mathbf{p}$ are collinear and related by the rational factor $\frac{2m+1}{2m'+1}$.

Lack of orthogonality can be tolerated for a basis set, but completeness is essential. This raises the question whether the extra $\delta$-functions spoil not only orthogonality but also completeness. Since (14) depends on (13) to prove completeness, a different method is needed for anharmonic waves.

Because sinusoidal waves are complete, one can demonstrate completeness of anharmonic waves by expanding the sinusoidal waves $\exp(iz)=\exp(-i\,\mathbf{p}\mathbf{x})$ into anharmonic waves. For Class 2 anharmonic waves this expansion takes the form:

(17) $\quad \exp(i\,z) = \sum_{m=-\infty}^{\infty} b_m \cdot \psi_2(m\,z) \qquad b_m = \sum_{n=0}^{\infty} b_{m,n}\, q^n \qquad m = -\infty,\ldots,\infty$

There is a well-defined procedure for obtaining the expansion coefficients $b_m$ after expanding them into power series of the anharmonicity parameter $q$ (see Appendix B). For $n=0$ one simply replaces $\exp(i\,z)$ with $\psi_2(z)$ by setting $b_{0,0}=1$. To proceed from $O(q^n)$ to $O(q^{n+1})$ one needs to eliminate every excess term of the form $q^{n+1}\exp(i\,m\,z)$ in (17) by adding the anharmonic wave of the type $q^{n+1}\psi_2(m\,z)$ with the opposite amplitude. That brings the accuracy of the expansion from $O(q^n)$ to $O(q^{n+1})$.

## 7. Anharmonic Quantum Fields

To define an anharmonic quantum field one can use standard methodology, where a quantum field operator $\Phi(x)$ is expanded into a series of sinusoidal plane waves multiplied by creation and annihilation operators. Such an expansion is motivated by the Fourier expansion of a classical field $\phi(x)$. The creation and annihilation operators correspond to the Fourier amplitudes. Such an expansion is always possible, because sinusoidal plane waves form a complete set, even if they are not solutions of the field equations. Since anharmonic waves are complete, too, they can be used as an alternative basis set. In particular, one can choose anharmonic solutions of the interacting classical field equations as basis. There is one caveat: A finite series of anharmonic waves is usually not a solution of the field equations, because the superposition principle breaks down for nonlinear field equations. An infinite series is required instead.

In the following, this concept will be demonstrated for a real scalar field $\Phi = \Phi^\dagger$. Analogous expansions for spinor and vector fields are presented in a separate publication [25]. In the Heisenberg picture, the space-time dependent quantum field operator $\Phi(x)$ can be defined by an expansion into sinusoidal plane waves, combined with creation and annihilation operators for particles with momentum $\mathbf{p}$ and energy $E = (\mathbf{p}^2 + M^2)^{1/2}$:

(18) $\boxed{\Phi(x) = (2\pi)^{-3/2} \int d^3p\, (2E)^{-1/2} \cdot \left[a(\mathbf{p}) \cdot e^{-ipx} + a^\dagger(\mathbf{p}) \cdot e^{+ipx}\right]}$ for $\Phi = \Phi^\dagger$

$a(\mathbf{p})$ = annihilation operator $\quad a^\dagger(\mathbf{p})$ = creation operator

$[a(\mathbf{p}), a^\dagger(\mathbf{p}')] = \delta^3(\mathbf{p}-\mathbf{p}'); \quad [a(\mathbf{p}), a(\mathbf{p}')] = [a^\dagger(\mathbf{p}), a^\dagger(\mathbf{p}')] = 0$

$x^\mu = (t, \mathbf{x}) \quad p^\mu = (p^0, \mathbf{p}) \quad E = |p^0| = (\mathbf{p}^2 + M^2)^{1/2} \quad p_\mu p^\mu = M^2 \quad p_\mu x^\mu = px$

The operator expansion (18) is generalized to anharmonic waves by replacing the sinusoidal plane wave $e^{-ipx}$ with its anharmonic counterpart $\psi(-px)$:

(19) $\boxed{e^{-ipx} \Rightarrow \psi(-px)} \quad \psi(-px) = \psi_2(-px) = \sum_{m=-\infty}^{\infty} a_m \cdot e^{-i(2m+1)px}$ for Class 2

Expanding $\psi_2(-px)$ into its Fourier series (4a) leads to additional substitutions for the harmonics:

(20) $p_\mu \Rightarrow (2m+1) \cdot p_\mu \qquad E \Rightarrow |2m+1| \cdot E \qquad M \Rightarrow |2m+1| \cdot M$

$\int d^3p\, (2E)^{-1/2} \Rightarrow \int d^3(|2m+1|\cdot \mathbf{p})\, (|2m+1|\cdot E)^{-1/2} = \int d^3p\, (2E)^{-1/2} \cdot |2m+1|^{5/2}$

The energy versus momentum relation stays $E = (\mathbf{p}^2 + M^2)^{1/2}$, only multiplied by $|2m+1|$.

The Fourier series in (19) brings us back to sinusoidal waves which can be handled according to standard procedures. The creation and annihilation operators for the harmonics are standard creation and annihilation operators for sinusoidal waves:

(21) $\quad a(\mathbf{p}) \Rightarrow \sum_{m=-\infty}^{\infty} a_m \cdot a((2m+1)\mathbf{p}) \qquad a^\dagger(\mathbf{p}) \Rightarrow \sum_{m=-\infty}^{\infty} a_m \cdot a^\dagger((2m+1)\mathbf{p})$

They obey commutator relations involving the momenta of the harmonics:

(22) $\quad [a((2m+1)\mathbf{p}), a^\dagger((2m'+1)\mathbf{p}')] = \delta^3((2m+1)\mathbf{p} - (2m'+1)\mathbf{p}')$

$\qquad [a((2m+1)\mathbf{p}), a((2m'+1)\mathbf{p}')] = [a^\dagger((2m+1)\mathbf{p}), a^\dagger((2m'+1)\mathbf{p}')] = 0$

The substitutions (19)-(21) define the anharmonic field operator $\Phi_2$ of Class 2 waves by its expansion into sinusoidal plane waves:

(23) $\quad \boxed{\Phi_2(x) = \sum_{m=-\infty}^{\infty} a_m \cdot (2\pi)^{-3/2} \int d^3p \, (2E)^{-1/2} \cdot |2m+1|^{5/2} \cdot \left[ a((2m+1)\mathbf{p}) \cdot e^{-i(2m+1)px} + a^\dagger((2m+1)\mathbf{p}) \cdot e^{+i(2m+1)px} \right]}$

This is the generalization of (18) to anharmonic waves. Each term in this Fourier series has the standard form (18), but with $p_\mu$ multiplied by $(2m+1)$. The corresponding mass $((2m+1)p_\mu \cdot (2m+1)p^\mu)^{1/2}$ becomes multiplied by $|2m+1|$. Thus one can interpret the Fourier series as a composite particle consisting of the fundamental with mass $M$, plus heavy copies with the masses $|2m+1|\cdot M$. The Fourier coefficients $a_m$ determine the weight of the harmonics in such a way that the combined probability density of all constituents remains 1, due to the sum rule (4b).

It is interesting to compare this expansion of a quantum field into into anharmonic plane waves with another such expansion, where the exact solutions of the Dirac equation for an electron moving in the field of an electromagnetic wave were used [26-28]. In that case the wave function has the form of a Fourier series describing energy and momentum transfer between the electron and an integer number of photons from the electromagnetic wave. There are significant differences, though. The Dirac equation of an electron moving in an external field remains linear. Furthermore, the external electromagnetic field imposes its own periodicity, which is different from the period of the electron wave function. As a result, the four-momentum of the fundamental is that of the electron, while the harmonics add multiples of the photon four-momentum that corresponds to the external electromagnetic wave.

More generally, there is a close connection between a quantum field $\Phi(x)$ and the corresponding basis set of plane waves $\phi_p(x)$:

(24) $\quad \Phi(x) = \sum_p \left( a(\mathbf{p}) \cdot \phi_p(x) + a^\dagger(\mathbf{p}) \cdot \phi_p^*(x) \right)$

The operators $a(\mathbf{p})$ and $a^\dagger(\mathbf{p})$ annihilate and create the single particle state $|p\rangle = |\phi_p\rangle$ which is characterized by its four-momentum $p$ and its wave function $\phi_p(x)$. In other words, the state $|\phi_p\rangle$ is produced by applying the creation operator $a^\dagger(\mathbf{p})$ to the vacuum state $|0\rangle$, and it is converted back into the vacuum state by applying $a(\mathbf{p})$:

(25) $\quad |\phi_p\rangle = a^\dagger(\mathbf{p})|0\rangle \qquad a(\mathbf{p})|\phi_p\rangle = |0\rangle$

The wave function $\phi_p(x)$ can be obtained as matrix element of the field operator $\Phi(x)$ between the single-particle state $|\phi_p\rangle$ and the vacuum state $\langle 0|$:

(26) $\quad \phi_p(x) = \langle 0|\Phi(x)|\phi_p\rangle$

For sinusoidal waves one obtains the exponential with a normalization factor:

(27) $\quad \phi_{s,p}(x) = (2\pi)^{-3/2} \cdot e^{-ipx}$

Anharmonic waves of Class 2 have the normalized wave functions:

(28) $\quad \phi_{a,p}(x) = (2\pi)^{-3/2} \cdot \psi_2(-px) = \sum_m a_m \cdot \phi_{s,(2m+1)p}(x)$

This leads to an expansion of the creation operator $a^\dagger_a(\mathbf{p})$ for the anharmonic state $|\phi_{a,p}\rangle$ into sinusoidal creation operators $a^\dagger_s(\mathbf{p})$:

(29) $\quad a^\dagger_a(\mathbf{p})|0\rangle = |\phi_{a,p}\rangle = \sum_m a_m \cdot |\phi_{s,(2m+1)p}\rangle = \sum_m a_m \cdot a^\dagger_s((2m+1)\mathbf{p})|0\rangle$

$\Rightarrow \qquad a^\dagger_a(\mathbf{p}) = \sum_m a_m \cdot a^\dagger_s((2m+1)\mathbf{p})$

$\qquad\qquad a_a(\mathbf{p}) = \sum_m a_m \cdot a_s((2m+1)\mathbf{p})$

The particle number operator $a^\dagger(\mathbf{p})\cdot a(\mathbf{p})$ for a single particle state $|\phi_p\rangle$ is a fundamental quantity in quantum field theory. From (25) one can see that this operator indeed has 1 as eigenvalue for a single-particle state:

(30) $\quad a^\dagger(\mathbf{p})\cdot a(\mathbf{p})|\phi_p\rangle = a^\dagger(\mathbf{p})|0\rangle = 1 \cdot |\phi_p\rangle$

The same relation is less trivial for anharmonic waves, because a particle is broken up into its harmonics with non-integer Fourier coefficients. As self-consistency check one can verify (30) explicitly by expanding the anharmonic single particle states and operators according to (29):

$$\text{(31)} \quad \begin{aligned} a_a^\dagger(\mathbf{p}) \cdot a_a(\mathbf{p}) |\phi_{a,p}\rangle &= \sum_{m,m',m''} a_m\, a_{m'}\, a_{m''} \cdot a_s^\dagger((2m+1)\mathbf{p}) \cdot a_s((2m'+1)\mathbf{p}) |\phi_{s,(2m''+1)p}\rangle \\ &= \sum_{m,m',m''} a_m\, a_{m'}\, a_{m''} \cdot a_s^\dagger((2m+1)\mathbf{p}) \cdot \delta_{m',m''} \cdot |0\rangle \\ &= \sum_{m,m'} a_m \cdot a_{m'}^2 \cdot |\phi_{s,(2m+1)p}\rangle \\ &= \sum_{m'} a_{m'}^2 \cdot |\phi_{a,p}\rangle \\ &= 1 \cdot |\phi_{a,p}\rangle \end{aligned}$$

The factor $\delta_{m',m''}$ is due to the annihilation operator, which produces the zero-particle vacuum state $|0\rangle$ for $m'=m''$ and the number 0 for $m'\neq m''$. It eliminates the sum over $m''$ and creates a sum over $a_{m'}^2$. The vacuum state is then converted back to a series of sinusoidal single particle states by a series of creation operators. This is identical to the expansion of the anharmonic state into sinusoidal states in (29). The remaining sum over $m'$ gives 1 according to the sum rule (4b). Since (4b) resulted from the condition (2) $\psi^*\psi=1$, this basic requirement for classical fields ensures an integer particle number for quantum fields.

## 8. Conclusions and Outlook

In summary, sinusoidal plane waves are generalized to anharmonic waves for obtaining an *ansatz* for the exact field operators and single particle states of interacting quantum field theories. Criteria are given that make anharmonic waves compatible with relativistic field theory and quantum physics. They can be satisfied by multiplying a sinusoidal wave with a periodic phase factor. The transition from classical to quantum fields is accomplished by using anharmonic waves in the expansion of the field operators.

The next step will be the development of Feynman rules for anharmonic quantum fields. These are needed for a comparison with well-tested quantum field theories, such as quantum electrodynamics. After that, one can evaluate Feynman diagrams describing the generation of harmonics and thereby determine the Fourier coefficients of the harmonics as function of the coupling constant $\alpha$. This concept will be fleshed out for quantum electrodynamics in a subsequent publication [25]. Further down the road one could imagine calculating $\alpha$ itself by inserting exact field operators into non-perturbative relations, such as truncated versions of the Dyson-Schwinger equations [14],[29].

**Appendix A: Fourier Coefficients and Anharmonicity Parameter**

For Class 2 anharmonic waves the Fourier series (4a) takes the form:

(A1) $\quad \psi_2(z) = \sum_{-\infty}^{\infty} a_m \cdot \exp[i(2m+1)z]$

The values of $\psi_2(z)$ at the symmetry points $z=0$, $\frac{1}{2}\pi$, and $\pi$ in Table 1 generate the following restrictions for the coefficients $a_m$:

(A2) $\quad \psi_2(0) = 1 \qquad \psi_2(\pm\pi) = -1 \quad \Leftrightarrow \quad \boxed{\sum_{-\infty}^{\infty} a_m = 1}$

$\qquad\qquad\qquad \psi_2(\pm\pi/2) = \pm i \qquad\qquad \Leftrightarrow \quad \boxed{\sum_{-\infty}^{\infty} (-)^m a_m = 1}$

Adding and subtracting these conditions one finds:

(A3) $\quad \sum_{-\infty}^{\infty} a_{2m} = 1 \qquad \sum_{-\infty}^{\infty} a_{2m+1} = 0$

The condition $\psi^*\psi = 1$ imposes an infinite number of constraints onto the coefficients of the product series:

(A4) $\quad \left(\sum_{-\infty}^{\infty} a_m \cdot \exp(-i\,2m\,z)\right) \cdot \left(\sum_{-\infty}^{\infty} a_n \cdot \exp(+i\,2n\,z)\right) = 1$

(A5) $\quad \boxed{\sum_{m=-\infty}^{\infty} a_m\, a_{m+n} = 0} \qquad$ for $\quad |n| = 1,\ldots,\infty$

(A6) $\quad \boxed{\sum_{m=-\infty}^{\infty} a_m^2 = 1} \qquad$ for $\quad n = 0$

Rather than dealing with an infinite number of constraints for the regular Fourier coefficients $a_m$, it is advantageous to start with the unconstrained Fourier coefficients $c_m$ in (5a,b) as input:

(A7) $\quad \psi_2(z) = \exp(i\,z) \cdot \exp\!\left[i\,2\sum_{m=1}^{\infty} c_m \cdot \sin(2mz)\right]$

This series can be truncated at any point, while the regular Fourier series (A1) violates condition (2) when truncated.

For controlling the truncation error one can expand the Fourier coefficients $a_m$, $c_m$ into Taylor series of a parameter $q$ that characterizes the strength of the anharmonicity:

(A8) $\quad \boxed{a_m = \sum_{n=|m|}^{\infty} a_{m,n}\, q^n} \qquad m = -\infty,\ldots,\infty$

(A9) $\quad \boxed{c_m = \sum_{n=m}^{\infty} c_{m,n}\, q^n} \qquad m = 1,\ldots,\infty$

The anharmonicity parameter q is modeled after the Nome q in the Fourier series of elliptic anharmonic waves (see Appendix D and [30]). A natural choice for q is the lowest Fourier coefficient $c_1$, which dominates the anharmonicity:

(A10) $\quad q = c_1 \quad \Rightarrow \quad c_{1,1} = 1 \quad c_{1,m} = 0 \quad$ for $m = 2, \ldots, \infty$

One can also make other choices $q'$, which include a scaling factor $c'_{1,1} \neq 1$ or non-vanishing higher order terms $c'_{1,m}$ (see Appendices C, D, E). They can be mapped back onto (A10) by the function $q'(q)$, which is the inverse of the function $q(q') = c_1(q')$.

The expansions (A8), (A9) can be used to convert the coefficients $c_m$ into $a_m$ or vice versa. Usually, one wants to start with the $c_m$ series, because it fulfills the condition (2) automatically, even when truncated. This is accomplished by inserting (A8) into (A1), (A9) into (A7), and expanding both into a power series of q. For each power $q^n$ the coefficients of all harmonics are matched, starting with $a_{0,0}=1$ for n=0 and proceeding towards higher n. With a complete set of $a_{m,n}$ and $c_{m,n}$ up to the power $q^n$ in hand, the coefficient $c_{m,n+1}$ is used as extra input to derive the coefficients $a_{m,n+1}$ for all m. This program is carried out in *Mathematica* 7, like many other the results in the Appendices. The choice $q=c_1$ in (A10) produces the following coefficients for Class 2 waves when truncated at $O(q^2)$:

(A11) $\quad\boxed{\begin{array}{llll} c_1 = q & a_0 \approx 1 - q^2 & a_{\pm 1} \approx \pm q & a_{\pm 2} \approx (½ \pm c_{2,2}) q^2 \\ c_2 \approx c_{2,2} q^2 & & & \end{array}}$

With the extra symmetry (6) one can go to $O(q^4)$ with the same number of parameters:

(A12) $\quad\boxed{\begin{array}{llll} c_{2m} = 0 & a_{-m} = (-)^m \cdot a_m & & \\ c_1 = q & a_0 \approx 1 - q^2 + ¼ q^4 & a_{\pm 1} \approx \pm(q - ½ q^3) & a_{\pm 2} \approx ½ q^2 - (⅙ + c_{3,3}) q^4 \\ c_3 \approx c_{3,3} q^3 & & a_{\pm 3} \approx \pm(⅙ + c_{3,3}) q^3 & a_{\pm 4} \approx (1/24 + c_{3,3}) q^4 \end{array}}$

It is advantageous to minimize the number of free parameters by choosing an even number n for the cutoff, since the even coefficients $c_{2m}$ vanish. New parameters appear with every odd power of q, starting with $c_{3,3}$ at $O(q^3)$, continuing with $c_{3,5}, c_{5,5}$ at $O(q^5)$, and escalating to $c_{3,7}, c_{5,7}, c_{7,7}$ at $O(q^7)$. Although short, the truncated series (A11),(A12) provide rather accurate approximations for an interaction with a small coupling constant $\alpha$, such as quantum electrodynamics. The anharmonicity depends on the interaction term in the Lagrangian, which is proportional to $\alpha$.

**Appendix B: Expansion of Sinusoidal Waves into Anharmonic Waves**

In order to demonstrate the completeness of anharmonic waves it is sufficient to expand the sinusoidal wave $\exp(iz)$ into a series anharmonic waves of the type $\psi(mz)$. Class 2 anharmonic waves with the extra symmetry (6) serve as example:

(B1) $\quad \psi_2(z) = \sum_{m=-\infty}^{\infty} a_m \cdot \exp[i(2m+1)z] \qquad a_m = \sum_{n=|m|}^{\infty} a_{m,n} q^n \qquad a_{-m} = (-)^m \cdot a_m$

(B2) $\quad \exp(iz) = \sum_{m=-\infty}^{\infty} b_m \cdot \psi_2[(2m+1)z] \qquad b_m = \sum_{n=0}^{\infty} b_{m,n} q^n$

(B2) is the inverse of the Fourier series (B1). While (B1) expands an anharmonic wave into sinusoidal waves, (B2) expands a sinusoidal wave into anharmonic waves.

To obtain the coefficients $b_m$ from the $a_m$, both are expanded into power series of q. Equal harmonics are compared for each power of q. After matching all Fourier coefficients up to $O(q^n)$, one encounters unwanted extra harmonics in $O(q^{n+1})$. These are eliminated by counter-terms of the form $b_{m,n+1} \cdot \psi_2[(2m+1)z] \cdot q^{n+1}$. The result simplifies greatly after taking care of the additional constraints (A5),(A6) for the $a_{m,n}$ by using the $c_{m,n}$ in (A9) parameters. In $O(q^4)$ one obtains for the coefficients $b_m$:

(B3) $\quad b_0 = 1 + 2q^2 + 15/4 \, q^4$

$b_{+1} = -q - 1\frac{1}{2}q^3 \qquad\qquad b_{+8} = 0$
$b_{-1} = +q + \frac{5}{2} q^3 \qquad\qquad b_{-8} = (19/6 - 2c_{33})q^4$

$b_{+2} = -\frac{1}{2} q^2 + (-2 + 3c_{33})q^4 \qquad \ldots$
$b_{-2} = -\frac{5}{2} q^2 + (-3\frac{1}{3} + c_{33})q^4 \qquad b_{+10} = (\frac{1}{3} + 2c_{33})q^4$

$b_{+3} = (-\frac{1}{6} - c_{33})q^3 \qquad\qquad \ldots$
$b_{-3} = (-\frac{5}{6} + c_{33})q^3 \qquad\qquad b_{+12} = +\frac{1}{4} q^4$

$b_{+4} = q^2 + (269/24 - c_{33})q^4 \qquad b_{+13} = -q^3$
$b_{-4} = \phantom{q^2 +}(-\frac{3}{8} - 3c_{33})q^4 \qquad \ldots$

$b_{+5} = 0 \qquad\qquad b_{-14} = -1\frac{1}{2}q^4$
$b_{-5} = 4q^3 \qquad\qquad \ldots$

$\ldots \qquad\qquad b_{+22} = -\frac{3}{2} q^4$

$b_{+7} = q^3 \qquad\qquad \ldots$
$b_{-7} = 0 \qquad\qquad b_{+40} = q^4$

This is the inverse of (A12). The dots stand for omitted coefficients that vanish. Although this expansion can be carried *ad infinitum*, it is rather inefficient. The coefficients $b_{m,n}$ of $O(q^n)$ generate harmonics up to $(2m+1) = 3^n$. For example, the highest coefficient in $O(q^4)$ is $b_{+40}$, which generates the 81st harmonic (n=4, m=40, (2m+1) = $3^n$ = 81).

**Appendix C: Simple Anharmonic Waves**

Anharmonic waves with explicit Fourier coefficients $a_m$ can be constructed by using only the first coefficient $c_1$ in the series (5a,b). They satisfy the extra symmetry (6):

(C1) $\psi_1(z) = \exp[i(z + q \cdot \sin(z))]$    $c_1 = \tfrac{1}{2} q$    Class 1

(C2) $\psi_2(z) = \exp[i(z + \tfrac{1}{2} q \sin(2z))]$    $c_1 = \tfrac{1}{4} q$    Class 2

(C3) $\psi_3(z) = \exp[i(z + \tfrac{1}{4} q \sin(4z))]$    $c_1 = \tfrac{1}{8} q$    Class 3

The corresponding coefficients $a_m$ are given by the oscillatory Bessel functions $J_m$:

(C4) $\psi_1(z) = \sum_{m=-\infty}^{\infty} a_m \cdot e^{+i(m+1)z}$    $a_m = J_m(q) \to (\tfrac{1}{2} q)^m / m!$   $q \to 0$

(C5) $\psi_2(z) = \sum_{m=-\infty}^{\infty} a_m \cdot e^{+i(2m+1)z}$    $a_m = J_m(\tfrac{1}{2} q) \to (\tfrac{1}{4} q)^m / m!$   $q \to 0$

(C6) $\psi_3(z) = \sum_{m=-\infty}^{\infty} a_m \cdot e^{+i(4m+1)z}$    $a_m = J_m(\tfrac{1}{4} q) \to (\tfrac{1}{8} q)^m / m!$   $q \to 0$

$J_0(q) = \sum_{n=0}^{\infty} (-\tfrac{1}{4} q^2)^n / (n!)^2 \approx 1 - \tfrac{1}{4} q^2$    $J_{-m} = (-)^m \cdot J_m$ $\Rightarrow$ $a_{-m} = (-)^m \cdot a_m$

The resulting anharmonic wave functions are plotted in Fig. 4 for the range $-1 \le q \le +1$. Beyond this range they exhibit additional oscillations which are unphysical. The anharmonicity parameter q is defined such that the range of physical q-values coincides with that of the elliptic functions in Appendix D.

Most of the characteristics of the three classes of anharmonic waves have already been discussed in Section 4 and Table 1. Figure 4 illustrates one extra symmetry. i.e., the behavior with the sign change of q. Here we need q explicitly as second variable of $\psi$:

(C7) Class 1:    $\psi_1(-q, z) = -\psi_1(q, z+\pi)$

Class 2:    $\text{Re}[\psi_2(-q, z)] = \text{Im}[\psi_2(q, z+\pi/2)]$    $\text{Im}[\psi_2(-q, z)] = -\text{Re}[\psi_2(q, z+\pi/2)]$

Class 1 functions exhibit a phase shift, but the real and imaginary part keep their shape. For Class 2 functions the shapes of the real and imaginary part are interchanged. Class 3 functions change their shape when changing the sign of q. This behavior is independent of the extra symmetry condition (6).

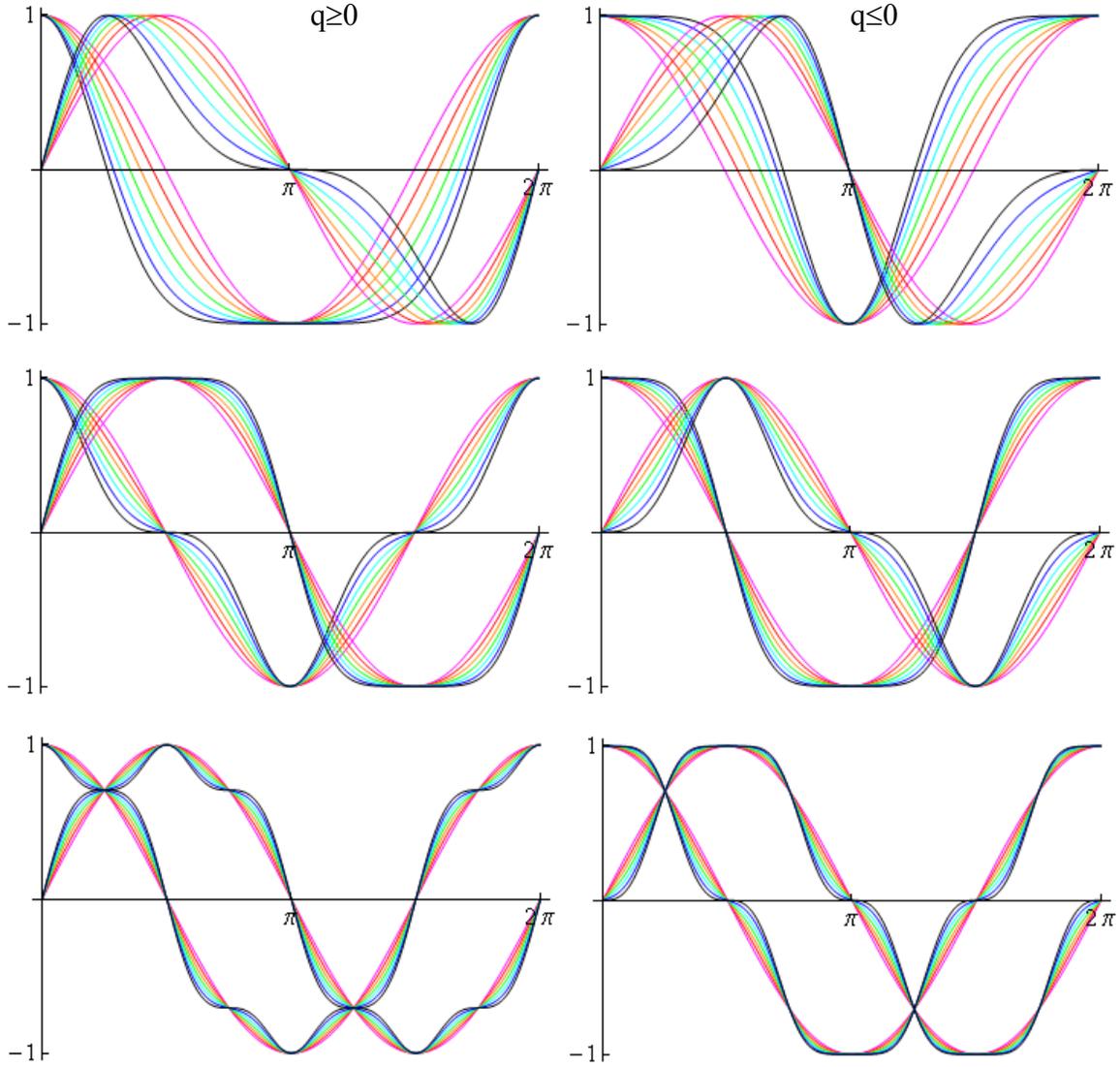

**Figure 4** Anharmonic waves in Class 1, 2, 3 (top to bottom) with the extra symmetry (6). The anharmonicity parameter q ranges from 0 to +1 on the left and from 0 to −1 on the right (magenta to black). Real and imaginary part are shown together. From (C1)-(C3).

So far we have considered only the lowest term $c_1$ in the phase series (5a,b). It dominates in field theories with weak coupling, such as quantum electrodynamics. Nevertheless, one might expect an infinite series of $c_m$ from pertubation theory. Therefore we also consider the infinite series $c_m = q^m/m$, which approximates the Fourier coefficients of the elliptic functions (D6)-(D8) for small q and large m. Since the even coefficients $c_{2m}$ do not vanish, these functions lack the extra symmetry (6). This series can be summed analytically to obtain the wave function $\psi$ in closed form:

(C8) $\varphi_1(z) = \sum_{m=1}^{\infty} q^m/m \cdot \sin(mz)$   $c_1 = \tfrac{1}{2} q$   Class 1

$\psi_1(z) = e^{iz} \cdot [(1 - q e^{-iz})/(1 - q e^{iz})]^{1/2}$

(C9) $\varphi_2(z) = \tfrac{1}{2} \cdot \sum_{m=1}^{\infty} q^m/m \cdot \sin(2mz)$   $c_1 = \tfrac{1}{4} q$   Class 2

$\psi_2(z) = e^{iz} \cdot [(1 - q e^{-i2z})/(1 - q e^{i2z})]^{1/4}$

(C10) $\varphi_3(z) = \tfrac{1}{4} \cdot \sum_{m=1}^{\infty} q^m/m \cdot \sin(4mz)$   $c_1 = \tfrac{1}{8} q$   Class 3

$\psi_3(z) = e^{iz} \cdot [(1 - q e^{-i4z})/(1 - q e^{i4z})]^{1/8}$

**Appendix D: Anharmonic Waves from Elliptic Functions**

Elliptic functions are particularly useful for constructing anharmonic waves. They are periodic along both the real and imaginary axis and converge to standard trigonometric functions when the anharmonicity vanishes (see [30], Ch. 16). They are also solutions of simple nonlinear differential equations and thus repeatedly appear in the soliton literature [19],[21]. Jacobi's elliptic functions form anharmonic waves of Class 2 after re-scaling the variable z to obtain the correct period $2\pi$. The functions sn and cd are "top-heavy", while cn and ($m_1^{1/2}$ sd) are "bottom-heavy":

(D1) $\psi_2(q,z) = \text{cn}(q, z') + i \, \text{sn}(q, z')$   $z' = z \cdot \tfrac{2K}{\pi}$

$\psi_2(-q,z) = \text{cd}(q, z') + i \, m_1^{1/2} \text{sd}(q, z')$   $-1 < q < 1$

(D2) $\text{cn}(-q, z') = \text{cd}(q, z')$

$\text{sn}(-q, z') = m_1^{1/2} \text{sd}(q, z')$

The Nome q is a parameter characterizing the anharmonicity of elliptic functions (see [30], 16.23 and 17.3.17). For q=0 one obtains sinusoidal functions. In fact, the anharmonicity parameter q is motivated by the Nome. Elliptic functions are usually defined for the range $0 \le q \le 1$. Here we extend their range to negative q-values by using the relation (D2) between elliptic functions with opposite signs of q. This relation will be established by generalizing their Fourier series to negative q in (D4).

A second anharmonicity parameter for elliptic functions is *m*. It determines the quarter-period K along the real axis, while $m_1 = 1 - m$ determines the quarter-period K' along the imaginary axis. K and K′ are elliptic integrals of $m$. Here we use q as independent parameter and express all other parameters in terms of q, including $m$.

(D3) $\quad m(q) = 16q - 128q^2 + 704q^3 - 3072q^4 + \ldots \qquad 0 \le m \le 1$

$$K = K[m(q)] = \pi/2 + 2\pi \cdot \sum_{n=1}^{\infty} q^n/(1+q^{2n}) \qquad \text{[30] 17.3.22}$$

$$= \pi/2 \cdot \left(1 + 2 \cdot \sum_{n=1}^{\infty} q^{n^2}\right)^2 \qquad \text{[30] 16.38.5, 16.27.3}$$

$$K'(m) = K(1-m) = K(m_1) \qquad m_1 = 1-m$$

$$q(m) = e^{-\pi \cdot K'(m)/K(m)} \qquad \text{[30] 16.23}$$

$$= (m/16) + 8\,(m/16)^2 + 84\,(m/16)^3 + 992\,(m/16)^4 + \ldots \qquad \text{[30] 17.3.21}$$

The period 4K becomes larger than $2\pi$ with increasing q. This stretch is compensated in (D1) by scaling the variable z to z′. Along the imaginary axis the elliptic exponential function $[cn(iz) + i\,sn(iz)]$ becomes $[nc(z) - sc(z)]$, which contains poles. For q<0 one has to consider $[cd(iz) + im_1^{1/2} sd(iz)]$, which becomes $[nd(z) - m^{1/2} sd(z)]$ along the imaginary axis. It remains oscillatory.

The Fourier series of elliptic functions from [30], 16.23 can be generalized to q<0 by extracting a factor $q^{1/2}$ from the coefficients and including it in the amplitude A(q). To avoid confusion with $m, m_1$ in (D3) the index m of the Fourier series is changed to n:

(D4) 
$$\begin{aligned}
cn(z) &= A(q) \cdot \sum_{n=0}^{\infty} (+q)^n/(1+q^{2n+1}) \cdot \cos[(2n+1)\,z''] & z'' &= z \cdot \tfrac{\pi}{2K} \\
sn(z) &= A(q) \cdot \sum_{n=0}^{\infty} (+q)^n/(1-q^{2n+1}) \cdot \sin[(2n+1)\,z''] & -1 &< q < 1 \\
cd(z) &= A(q) \cdot \sum_{n=0}^{\infty} (-q)^n/(1-q^{2n+1}) \cdot \cos[(2n+1)\,z''] \\
m_1^{1/2}\,sd(z) &= A(q) \cdot \sum_{n=0}^{\infty} (-q)^n/(1+q^{2n+1}) \cdot \sin[(2n+1)\,z'']
\end{aligned}$$

$$A(q) = 2\pi\,|q|^{1/2} / [\,K[m(|q|)] \cdot m(|q|)^{1/2}\,] = \left(\sum_{n=0}^{\infty} q^{n(n+1)}\right)^{-2} \qquad \text{[30] 16.38.2}$$

$$= 1 - 2q^2 + 3q^4 - 6q^6 + 11q^8 - 18q^{10} + \ldots \qquad \text{using } \vartheta_c(0) = 1$$

Changing the sign of q switches $cn \leftrightarrow cd$ and $sn \leftrightarrow (m_1^{1/2} sd)$ as spelled out in (D2). The Fourier series of the anharmonic exponential defined in (D1) is be obtained by adding the series for cn and $i$sn in (D4) and reshuffling the indices:

(D5) $\quad \boxed{\psi_2(z) = A(q) \cdot \sum_{n=-\infty}^{\infty} q^n/(1-q^{4n+2}) \cdot e^{+i(2n+1)z}} \qquad -1 < q < 1$

This expansion corresponds to the Fourier series (4a). The extra symmetry (6) does not hold, because $|a_{-n}| \neq |a_n|$ due to the sign change in the exponent of $q^n$ and $q^{4n+2}$.

The corresponding Fourier series (5a,b) for elliptic functions is obtained by converting the coefficients $a_n$ in (D5) to $c_n$ using the power series expansions (A8),(A9). These are then extrapolated to an infinite series in q. In addition to Class 2 waves one can define Class 1 and Class 3 waves by simply changing the period of the sine function:

(D6) $\quad \boxed{\varphi_1(z) = 2 \sum_{n=1}^{\infty} c_n \sin(n\,z) \qquad c_n = 2\,q^n/n \cdot (1+q^{2n})^{-1} \qquad c_1 \to 2q \quad q \to 0}$

(D7) $\quad \boxed{\varphi_2(z) = 2 \sum_{n=1}^{\infty} c_n \sin(2n\,z) \qquad c_n = \phantom{2}q^n/n \cdot (1+q^{2n})^{-1} \qquad c_1 \to q \quad q \to 0}$

(D8) $\quad \boxed{\varphi_3(z) = 2 \sum_{n=1}^{\infty} c_n \sin(4n\,z) \qquad c_n = \tfrac{1}{2}\,q^n/n \cdot (1+q^{2n})^{-1} \qquad c_1 \to \tfrac{1}{2}q \quad q \to 0}$

The ellipticity remains to be investigated for the Class 1 and Class 3 functions.

Neville's $\vartheta$-functions have Fourier expansions similar to those of Jacobi's functions, including the Nome q (see [30] 16.28, 16.38). However, they are not elliptic. Since the canonical $\vartheta$-functions $\vartheta_c$ and $m_1^{1/4}\vartheta_s$ have the same shape, two additional $\vartheta$-functions with a different shape are required to establish analogies to cn, sn, cd, $(m_1^{1/2} sd)$:

(D9) $\quad \boxed{\begin{aligned}
\vartheta_{cn}(z) &= B(+q) \cdot \sum_{n=0}^{\infty} \text{sign}(+q)^n \cdot q^{n(n+1)} \cdot \cos[(2n+1)\,z''] &&= \vartheta_c(q,z) \\
\vartheta_{sn}(z) &= B(-q) \cdot \sum_{n=0}^{\infty} \text{sign}(+q)^n \cdot q^{n(n+1)} \cdot \sin[(2n+1)\,z''] \\
\vartheta_{cd}(z) &= B(-q) \cdot \sum_{n=0}^{\infty} \text{sign}(-q)^n \cdot q^{n(n+1)} \cdot \cos[(2n+1)\,z''] \\
m_1^{1/4}\,\vartheta_{sd}(z) &= B(+q) \cdot \sum_{n=0}^{\infty} \text{sign}(-q)^n \cdot q^{n(n+1)} \cdot \sin[(2n+1)\,z''] &&= m_1^{1/4}\,\vartheta_s(q,z)
\end{aligned}}$

$\quad B(q) = \Big( \sum_{n=0}^{\infty} \text{sign}(q)^n \cdot q^{n(n+1)} \Big)^{-1}$

The four $\vartheta$-functions have been labeled analogous to their elliptic counterparts. The amplitude B(q) resembles the square root of the amplitude A(q) for elliptic functions in (D4), but the series for B(q) contains an extra factor $sign(q)^n$.

Unlike Jacobi's functions, $\vartheta$-functions do not satisfy the normalization condition $\psi^*\psi = 1$. There is a general method of converting a pair of oscillating functions $\vartheta_{cn}$ and $\vartheta_{sn}$ with cosine and sine characteristics into a properly-normalized anharmonic wave $\psi$:

(D10)  $Re[\psi] = \vartheta_{cn} / (\vartheta_{cn}^2 + \vartheta_{sn}^2)^{1/2}$

$Im[\psi] = \vartheta_{sn} / (\vartheta_{cn}^2 + \vartheta_{sn}^2)^{1/2}$

Following this recipe one can define Class 2 functions from the $\vartheta$-functions in (D9) by selecting pairs of top- and bottom-heavy functions. The two pairs correspond to opposite signs of q, similar to the elliptic functions in (D2).

(D11)  $Re[\psi_2(q,z)] = \vartheta_{cn}(q,z') / [\vartheta_{cn}(q,z')^2 + \vartheta_{sn}(q,z')^2]^{1/2}$

$Im[\psi_2(q,z)] = \vartheta_{sn}(q,z') / [\vartheta_{cn}(q,z')^2 + \vartheta_{sn}(q,z')^2]^{1/2}$

$Re[\psi_2(-q,z)] = \vartheta_{cd}(q,z') / [\vartheta_{cd}(q,z')^2 + m_1^{1/2}\vartheta_{sd}(q,z')^2]^{1/2}$

$Im[\psi_2(-q,z)] = m_1^{1/4}\vartheta_{sd}(q,z') / [\vartheta_{cd}(q,z')^2 + m_1^{1/2}\vartheta_{sd}(q,z')^2]^{1/2}$

**Appendix E:   Anharmonic Waves from Differential Equations and Lagrangians**

For establishing a connection between anharmonic waves and nonlinear differential equations it is natural to start with elliptic functions, which satisfy a variety of linear and nonlinear differential equations. Using the relations for the derivatives of elliptic functions in [30] 16.16 one finds the following first-order differential equations for the anharmonic cosine, sine, and exponential functions defined in (D1):

(E1)   $cn' = -sn \cdot dn$
       $sn' = +cn \cdot dn$       $\}$   $[cn + i\,sn]' = i\,[cn + i\,sn] \cdot dn$

(E2)   $cd' = -(m_1^{1/2}sd) \cdot (m_1^{1/2}nd)$
       $(m_1^{1/2}sd)' = +cd \cdot (m_1^{1/2}nd)$   $\}$   $[cd + i\,(m_1^{1/2}sd)]' = i\,[cd + i\,(m_1^{1/2}sd)] \cdot (m_1^{1/2}nd)$

Elliptic functions satisfy a variety of nonlinear differential equations. These can be obtained by eliminating dn(z) and nd(z) from (E1) and (E2). First, one swaps $cn \leftrightarrow cd$ and $sn \leftrightarrow sd$ on the right side via [30] 16.3.1, 16.3.2. Then one uses [30] 16.9.1 and 16.9.2 to eliminate $dn^2$ and $nd^2$ in favor of $cn^2, sn^2$ and $cd^2, sd^2$:

(E3) $$\boxed{\begin{aligned} cn' &= -(m_1^{1/2} sd) \cdot [1 + m/m_1 \cdot cn^2] \cdot m_1^{1/2} \\ (m_1^{1/2} sd)' &= +cn \cdot [1 + m/m_1 \cdot (m_1^{1/2} sd)^2] \cdot m_1^{1/2} \end{aligned}}$$ } "bottom-heavy"

(E4) $$\boxed{\begin{aligned} cd' &= -sn \cdot [1 - m \cdot cd^2] \\ sn' &= +cd \cdot [1 - m \cdot sn^2] \end{aligned}}$$ } "top-heavy"

These can be transformed into differential equations containing a single elliptic function:

$$F = cn, (m_1^{1/2} sd) \quad \text{"bottom-heavy"}$$
$$G = cd, sn \quad \text{"top-heavy"}$$

(E5) $\boxed{F'^2 = m_1 + (m-m_1) \cdot F^2 - m \cdot F^4}$

(E6) $\boxed{G'^2 = 1 - (1+m) \cdot G^2 + m \cdot G^4}$

(E7) $\boxed{F'' = +(m-m_1) \cdot F - 2m \cdot F^3}$

(E8) $\boxed{G'' = -(1+m) \cdot G + 2m \cdot G^3}$

There are no analogous differential equations for the anharmonic exponential. Its real and imaginary part have different shapes and thus must obey different differential equations. They coincide only for $m=0$ when both become sinusoidal.

The derivatives (E1), (E2) can be generalized beyond elliptic functions by replacing the elliptic functions dn, $(m_1^{1/2} nd)$ with a more general function $(1+h)$ which oscillates around 1:

(E9) $\boxed{\begin{aligned} f' &= -g \cdot (1+h) \\ g' &= +f \cdot (1+h) \end{aligned}}$   $\begin{aligned} f(z) &= \text{Re}[\psi(z)] \\ g(z) &= \text{Im}[\psi(z)] \end{aligned}$

The solutions automatically satisfy the normalization condition (2) when combined with appropriate starting values $f(0)=1$ and $g(0)=0$. This can be seen by multiplying the first equation of (E9) with $f$, the second equation with $g$, and adding them up:

(E10) $(f^2+g^2)' = 2 \cdot (ff' + gg') = 2 \cdot (-fg + gf) \cdot (1+h) = 0$

$\Rightarrow f^2 + g^2 = 1$ if $f(0)=1$ $g(0)=0$

The function $h(z)$ has to match the period of $f(z)$ and $g(z)$, and it needs to be even at $z=0$. This leads to a Fourier cosine series for $h(z)$:

(E11) $\boxed{\begin{aligned} h_1(z) &= \sum_{n=1}^{\infty} d_n \cdot \cos(n\, z) \\ h_2(z) &= \sum_{n=1}^{\infty} d_n \cdot \cos(2n\, z) \\ h_3(z) &= \sum_{n=1}^{\infty} d_n \cdot \cos(4n\, z) \end{aligned}}$  Class 1

Class 2

Class 3

$$q = ¼\, d_1$$

The anharmonicity parameter q has been defined such that it mimics the Nome of the elliptic functions. Such linear differential equations in the presence of an oscillatory potential are reminiscent of the Dirac equation in the presence of an electromagnetic wave [26-28].

Nonlinear differential equations for anharmonic waves can be constructed from (E9), (E11) by replacing the external function h(z) with a combination of the functions f(z) and g(z). Here we restrict ourselves to the dominant n=1 term of the Fourier series of h(z) and simply replace cos(z) by f(z) for Class 1. For Class 2 and 3 we expand cos(2z) and cos(4z) into $\cos^2(z)$ and $\sin^2(z)$, which are then replaced by $f^2(z)$ and $g^2(z)$:

(E12)  $\cos(z) \cdot g$ $\quad\quad\quad\quad\quad\quad\quad\quad \to fg$ $\quad\quad\quad\quad\quad\quad$ for Class 1

$\quad\quad\quad\cos(z) \cdot f$ $\quad\quad\quad\quad\quad\quad\quad\quad \to f^2$

$\quad\quad\quad\cos(2z)\cdot g = [\cos^2(z)-\sin^2(z)]\cdot g \quad \to (f^2-g^2)\cdot g = +(g-2g^3) \quad$ for Class 2

$\quad\quad\quad\cos(2z)\cdot f = [\cos^2(z)-\sin^2(z)]\cdot f \quad \to (f^2-g^2)\cdot f = -(f-2f^3)$

$\quad\quad\quad\cos(4z)\cdot g = [\cos^2(2z)-\sin^2(2z)]\cdot g \to [(f^2-g^2)^2-(2fg)^2]\cdot g \quad$ for Class 3
$\quad\quad\quad\quad\quad\quad\quad\quad\quad\quad\quad\quad\quad\quad\quad\quad\quad = (g-8g^3+8g^5)$

$\quad\quad\quad\cos(4z)\cdot f = [\cos^2(2z)-\sin^2(2z)]\cdot f \to [(f^2-g^2)^2-(2fg)^2]\cdot f$
$\quad\quad\quad\quad\quad\quad\quad\quad\quad\quad\quad\quad\quad\quad\quad\quad\quad = (f-8f^3+8f^5)$

For Class 2 and 3 one can rewrite the equations to avoid mixed f, g products, using the condition $f^2+g^2 = 1$. That facilitates the construction of a Lagrangian. Inserting the substitutions (E12) into (E9),(E11) one obtains nonlinear differential equations between f and g that do not contain the external function h. These differential equations have Class 1,2,3 anharmonic waves as solutions for the initial conditions f(0)=1, g(0)=0:

(E13)
$$f' = -\kappa \cdot g - 4q \cdot fg \quad\quad\text{Class 1}$$
$$g' = +\kappa \cdot f + 4q \cdot f^2$$

$$f' = -\kappa \cdot g - 4q \cdot (g-2g^3) \quad\quad\text{Class 2}$$
$$g' = +\kappa \cdot f - 4q \cdot (f-2f^3)$$

$$f' = -\kappa \cdot g - 4q \cdot (g-8g^3+8g^5) \quad\quad\text{Class 3}$$
$$g' = +\kappa \cdot f + 4q \cdot (f-8f^3+8f^5)$$

(E14) $\kappa = [1+(4q)^2]^{1/2}$

The factor $\kappa$ adjusts the period to $2\pi$. One could use the same method to generate nonlinear differential equations for the higher members of the Fourier series of h(z). That would generate higher order polynomials of f and g in (E13).

It is possible to find Lagrangians $L$ that have the differential equations (E9) and (E13) as Euler-Lagrange equations:

$$[\partial L/\partial g']' = \partial L/\partial g$$
$$[\partial L/\partial f']' = \partial L/\partial f$$

The linear differential equations (E9) with the external function h(z) have the Lagrangian:

(E15) $L_h = \tfrac{1}{2}(f \cdot g' - g \cdot f') - \tfrac{1}{2}(f^2 + g^2) \cdot (1+h)$

The nonlinear differential equations (E13) have the following Lagrangians for Class 2,3:

(E16) 
$L_2 = \tfrac{1}{2}(f \cdot g' - g \cdot f') - \tfrac{1}{2}\kappa \cdot (f^2 + g^2) + 2q \cdot [(f^2 - g^2) - (f^4 - g^4)]$
$L_3 = \tfrac{1}{2}(f \cdot g' - g \cdot f') - \tfrac{1}{2}\kappa \cdot (f^2 + g^2) - 2q \cdot [(f^2 + g^2) - 4(f^4 + g^4) + 8/3(f^6 + g^6)]$

These Lagrangians contain the potentials:

(E17) $V_2 = \tfrac{1}{2}\kappa \cdot (f^2 + g^2) - 2q \cdot [(f^2 - g^2) - (f^4 - g^4)]$

$V_3 = \tfrac{1}{2}\kappa \cdot (f^2 + g^2) + 2q \cdot [(f^2 + g^2) - 4(f^4 + g^4) + 8/3(f^6 + g^6)]$

For Class 1 it is less straightforward to obtain a Lagrangian and a potential without introducing square roots. A simple Lagrangian would be desirable to simulate spontaneous symmetry breaking and a Higgs potential.